\documentclass[letterpaper, 10 pt, conference]{ieeeconf}  

\IEEEoverridecommandlockouts                              

\usepackage{algorithm}
\usepackage{amsmath}
\usepackage{color,soul}
\usepackage{units}

   \usepackage[pdftex]{graphicx}
   \graphicspath{{./Figures/},{./}}
   \DeclareGraphicsExtensions{.pdf,.jpeg,.png}

\title{A Two-layer Decentralized Control Architecture for DER Coordination}

\author{Thomas Navidi$^{1}$, Abbas El Gamal$^{1}$, and Ram Rajagopal$^{2}$
\thanks{*The work presented in this paper was partially supported under ARPA-E award DE-AR0000697.}
\thanks{$^{1}$Thomas Navidi and Abbas El Gamal are with the Department of Electrical Engineering
	Stanford University, Stanford, California
	{\tt\small tnavidi@stanford.edu}}%
\thanks{$^{2}$Ram Rajagopal is with the Department of Civil and Environmental Engineering, Stanford University}
}

\begin{document}

\maketitle

\thispagestyle{empty}
\pagestyle{empty}

\begin{abstract}
	This paper presents a two-layer distributed energy resource (DER) coordination architecture that allows for separate ownership of data, operates with data subjected to a large buffering delay, and employs a new measure of power quality. The two-layer architecture comprises a centralized model predictive controller (MPC) and several decentralized MPCs each operating independently with no direct communication between them and with infrequent communication with the centralized controller. The goal is to minimize a combination of total energy cost and a measure of power quality while obeying cyber-physical constraints. The global controller utilizes a fast AC optimal power flow (OPF) solver and extensive parallelization to scale the solution to large networks. Each local controller attempts to maximize arbitrage profit while following the load profile and constraints dictated by the global controller. Extensive simulations are performed for two distribution networks under a wide variety of possible storage and solar penetrations enabled by the controller speed. The simulations show that (i) the two-layer architecture can achieve tenfold improvement in power quality relative to no coordination, while capturing nearly all of the available arbitrage profit for a moderate amount of storage penetration, and (ii) both power quality and arbitrage profits are optimized when the solar and storage are distributed more widely over the network, hence it is more effective to install storage closer to the consumer.
\end{abstract}


%
\IEEEpeerreviewmaketitle

\section{Introduction}

A grand challenge of the future electric grid is how to best coordinate distributed energy resources (DERs), such as rooftop photovoltaics (PV), energy storage, and controllable loads, to achieve grid reliability, provide ancillary services, and minimize energy cost, while allowing renewables to ultimately replace fossil fuel sources. DERs are currently operating autonomously, resulting in grid reliability problems as demonstrated by Hawaii's end of solar net-metering~\cite{hawaii}. Such reliability problems limit DER penetration and reduce its potential benefits; as we will show, however, when appropriately coordinated, the improvements can be quite significant. This paper will specifically address the benefits of coordinating distributed energy storage (DES) across a distribution network focusing on optimizing shifting the load through energy arbitrage by charging during off-peak hours and discharging during peak hours, while reducing voltage violations caused by the rapidly increasing installation of rooftop PV and its projected growth into the future~\cite{DOE_REPORT}.

Coordinating thousands of connected devices in the real world, however, is challenging due to cyberphysical and data constraints. Several authors \cite{DMPC,tapMPC,Guo2017-fx} have recently proposed using a model predictive control (MPC) approach due to its ability to cope with uncertainties due to generation and load variations. A centralized MPC architecture that includes all of the network parameters would achieve high performance, but would be too large to be computationally feasible for real time control as discussed in \cite{DMPC}. Several recent papers address this issue by distributing the computation across nodes in the network \cite{DMPC,boyd_messaging} \cite{junjie,consensus,garcia2012} using message passing to achieve similar performance to a fully centralized control. These fully distributed control architectures, however, fail to address practical concerns surrounding device ownership, data privacy, and communication infrastructure. Utility companies own the network information and collect smart meter data, but cannot control or measure behind the meter devices such as DERs. By contrast, a DER provider has access to its installed devices, but does not know the network, customer loads, or the data associated with devices owned by other DER providers. This dichotomy creates unnecessary conflicts between the utilities and DER providers that a successful coordination architecture must address. Furthermore, privacy concerns prevent sending information between individual consumers, and the massive number of devices and data to be collected within a distribution network prevent reliable communication at all times. In particular, the smart meter infrastructure can create a buffering delay of several hours in communicating the data to where it is needed for coordination~\cite{smart_meter_issues}. Moreover, there is no existing infrastructure capable of supporting peer-to-peer communication with high reliability and low delay, and installing such infrastructure over distribution networks would not be economically viable. 

\begin{figure}[!t]
	\centering
	\includegraphics[width=2.5in]{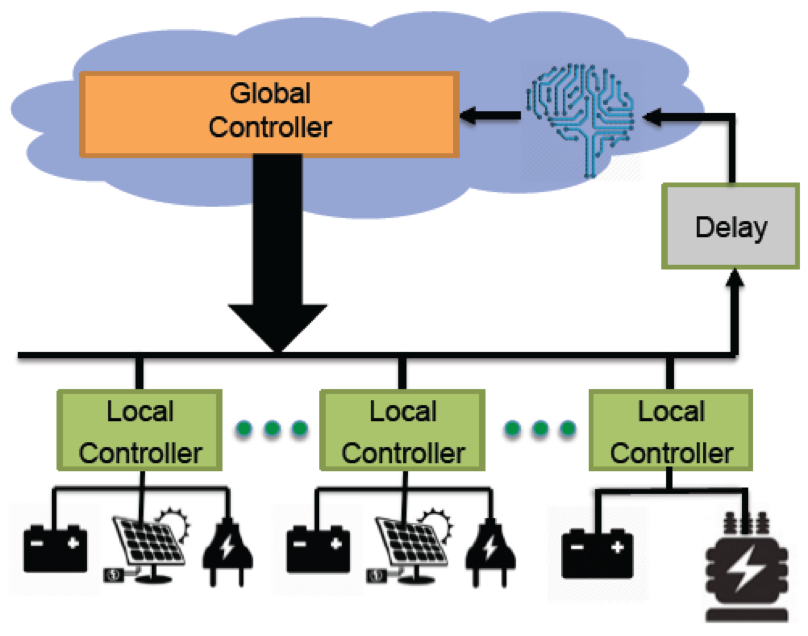}
	\caption{Illustration of the two-layer architecture.}
	\label{architecture}
\end{figure}

A two-layer architecture that incorporates aspects of both centralized and decentralized control in order to work around the aforementioned challenges is presented in~\cite{kyle}; see Figure~\ref{architecture}. This architecture distributes the coordination between a global controller (GC) and local controllers (LC) that all operate in an MPC fashion. Each storage system in the network is equipped with a local controller that has access to updated data only from the node it is connected to, and is unable to communicate with other nodes for privacy. The LCs communicate to the GC asynchronously and can have delays up to several hours. The GC is aware of the network model and makes decisions that aim to maintain reliability and minimize operational costs, while each LC uses its local data and the signals from the GC to maximize its own arbitrage profit. The GC could utilize cloud computing operated by the utility company, while the LCs can be run on the devices installed by the DER providers or by a third party. It was shown that the two-layer architecture is surprisingly robust to communication delays exceeding 24 hours due to the MPC and the ability of each LC to accurately forecast its future load and solar in the short term. Hence, this two-layer architecture combines the advantage of centralized control's minimal communication requirements with decentralized control's ability to run frequently due to low computational cost, without significant loss in performance.

This paper builds upon the coordination architecture in~\cite{kyle} in several respects. First, we incorporate a new electric power quality metric into the objective function, which enables us to quantify the adverse effects of excessive voltage deviations and enables the MPC to minimize these deviations without constraining the controller. Second, we introduce a new local control scheme that provides local flexibility without compromising global network reliability. Third, we significantly improve the computational efficiency of the architecture by using a faster optimal power flow (OPF) algorithm and extensive parallelization. This enables us to perform extensive simulations on two distribution networks under a wide variety of solar and storage configurations. We compare our two-layer architecture performance to a simple uncoordinated storage controller that achieves maximum arbitrage, but fails to address voltage violations in the network. We find that the two-layer controller can achieve around tenfold improvement in power quality relative to no coordination, while capturing most of the available arbitrage profit. This improvement occurs even at low levels of storage penetration relative to the solar penetration. Other simulations suggest that both power quality and arbitrage profits are optimized when the solar and storage are distributed more widely over the network, hence it is more effective to install storage spread across the consumer level than to aggregate it at the higher levels of a distribution network.

The rest of the paper is organized as follows. The next section introduces the models and assumptions used throughout the paper. Section~\ref{Architecture} provides a detailed description of the dual layer control scheme. Section~\ref{Results} describes our implementation and simulation results for two distribution networks.
\section{Models and assumptions} \label{NetworkModelSec}

We consider a radial distribution network consisting of $N$ nodes, $i\in \{0,1,2,\ldots,N\}=[0:N]$. The AC power flows in this network are discretized over $T$ time steps each of length $\delta_\mathrm{min}$. A subset of the nodes each has an aggregated collection of a stochastic uncontrollable load. A smaller subset of these nodes, $\mathcal{Z}$ also have solar and storage. We denote by $d_{i\tau}$ the sum of the uncontrollable loads (demand) and solar generation for node $i$ at time $\tau$. The nodes are connected via transmission lines represented as a set of edges $\mathcal{E}$ with admittances represented by the $Y_\mathrm{bus}$ matrix with entries $y_{iij}$. The complex power injected at each node $i$ is denoted by $s_{i\tau}$ and its real part by $\Re(s_{i\tau})$. The voltage at each node is denoted by $v_{i\tau}$. 

\noindent{\bf Storage constraints}. Storage units charge or discharge only real power. Each storage unit has a net charging rate of $u_{i\tau}$ at time $\tau$ and a state of charge of $q_{i\tau}$ with a maximum of $q^{\mathrm{max}}_i$. The storage dynamics are given by $q_{i\tau} = u_{i\tau}\cdot \delta_{\mathrm{min}} + \eta_i q_{i\tau-1}$ with a leakage coefficient of $0 \le \eta_i \le 1$. In a more practical implementation, we may also want to include more detailed dynamics such as charging and discharging efficiency or battery cycle life as in \cite{Anderson2016}. We do not include these constraints here for simplicity and because they are not essential to the main goals of this paper. Although assuming a perfectly efficient battery results in a pole at the origin, we have not observed any instability as a result of this. Hence, we have decided not to included charging and discharging loss in the formulation.

To ensure that a more complex battery model is indeed nonessential to the main points of this paper, we performed limited simulations using storage dynamics given by $q_{i\tau} = \mu_i u_{i\tau}\cdot \delta_{\mathrm{min}} - \nicefrac{v_{i\tau}}{\mu_i}\cdot \delta_{\mathrm{min}}+ \eta_i q_{i\tau-1}$, where $u_{i\tau} \ge 0$ and $v_{i\tau} \ge 0$ are the charging and discharging rate, respectively, and $0 \le \mu_i \le 1$ is the charging efficiency. While including these dynamics with less than perfect efficiency generally reduces the performance of the system, the controller remained numerically stable and performed with the same trends as using the simpler model.

\noindent{\bf Network constraints}. Denote the total net power demanded by each node as $\Re(s_{i\tau}) =  d_{i\tau} +  u_{i\tau}$. The voltage and power flows through the network are determined by the power flow constraints
\begin{equation} \label{PF}
s_{i\tau} = v_{i\tau} \sum_{j:(i,j)\in E} (v^{*}_{j\tau} - v^{*}_{i\tau})y^{*}_{ij}.
\end{equation}
This constraint is not convex due to the quadratic relationship between voltage and complex power. As described in~\cite{low_exact_relaxation}, however, it can be expressed as a second order cone convex relaxation.  This relaxation has been shown to be exact for many radial distribution networks, and is significantly faster than semidefinite programming based methods. The relaxation works by forming a new semidefinite rank 1 matrix $W = vv^T$, where $v$ is the node voltage vector. The rank 1 constraint on $W$ is relaxed, and the semidefinite constraint is broken up for each edge on the distribution network graph. This creates a series of 2x2 submatrices each containing the outer product of one node and one neighbor. The semidefinite constraint for a 2x2 matrix is a second order cone constraint. This constraint as well as the new power flow constraint for a single time step are as follows
\begin{subequations} \label{PFSOCP}
	\begin{alignat}{2}
	& s_{i} = \sum_{j:(i,j)\in E} (w_{ij} - w_{ii})y_{ij}^{*} \label{SOCP1} \\
	& \frac{w_{ij}^2 + w_{jj}^2}{w_{ii}} - w_{ij} \le 0 \hspace{3em} \forall (i,j) \in \mathcal{E} \label{SOCP2} \\
	& [w_{ii}]_{e^f} = [w_{jj}]_{e^t} \hspace{5em} \forall e \in \mathcal{E}. \label{SOCP3}
	\end{alignat} 
\end{subequations}
In equation \eqref{SOCP3}, $[w_{ii}]_{e^f}$ is the squared voltage of the ``from" node of edge $e^f$, and $[w_{jj}]_{e^t}$ is the ``to" node of adjacent edge $e^t$. This maintains consistency between submatricies and the original large outer product matrix. For simplicity, these semidefinite constraints for each submatrix is represented succinctly as $W\{i,j\}_\tau \succeq 0$ for time step $\tau$.

Now that the problem constraints can be formulated as a convex optimization problem, the goal then becomes how to determine $u_{i\tau}$ for all $\tau$ and all $i$ subject to the power flow, storage, and information constraints with the goals of minimizing operating costs and promoting network reliability. Note that the power flow equality constraints are always guaranteed by the substation node because it is allowed to generate as much real and reactive power from the transmission grid as needed to meet the demands of the nodes underneath it. Therefore, our control scheme has no role in ensuring the stability and satisfaction of this constraint. It is included in order to determine node voltages from their power injections.

\begin{figure}[!t]
	\centering
	\includegraphics[width=2.5in]{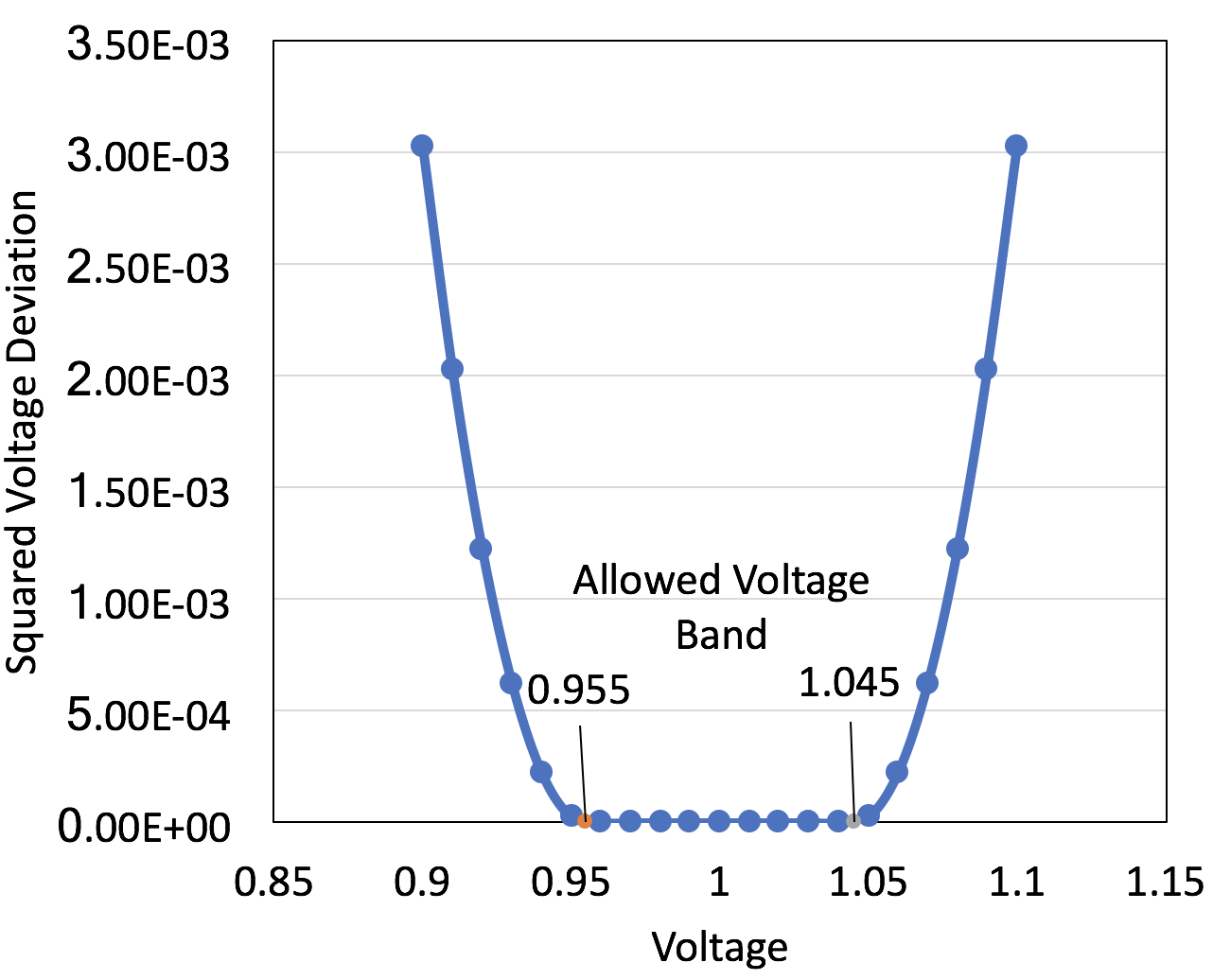}
	\caption{Lack of electric power quality metric measured as squared voltage deviations.}
	\label{vMetric}
\end{figure}

\noindent{\bf Power quality metric}. Our metric for network reliability is related to the electric power quality, particularly with respect to voltage violations. A lack of quality of service is quantified by the sum of squared voltage deviations above or below the nominal ($\pm 5$\%) bands 
\begin{equation}
\sum_{i=0}^{N}\sum_{t=1}^{T}(\max(v_{it} - 1.05, 0) + \max(0.95 - v_{it}, 0))^2, \label{Vmetric}
\end{equation}
where $v_{it}$ is the per unit voltage at node $i$ and time $t$. Figure \ref{vMetric} plots the squared voltage deviation as a function of voltage. This metric replaces the hard voltage constraints imposed in transmission system OPF. It helps eliminate the possible infeasibility of the solution caused by some voltage violations (that can otherwise be eliminated by voltage regulators not included in our model), and at the same time curb significant voltage deviations due to the introduction of RDGs. This band can be made 0.5\% tighter than the nominal bands in order to have sufficiently high penalty when used as a the cost function as shown in figure \ref{vMetric}.

\noindent{\bf Cost function}. After adjusting voltage variables to fit the convex relaxation of the OPF, the total cost function including the cost of electricity becomes
\begin{multline}
\sum_{\tau}   p_{\tau} \cdot \Re(s_{0\tau}) + \\ 
\lambda \sum_{i=0}^{N}\sum_{\tau}(\max(w_{ii\tau} - V_\mathrm{tol+}^2, 0) + \max(V_\mathrm{tol-}^2 - w_{ii\tau}, 0))^2,
\end{multline}
where $\lambda \ge 0$ is a parameter that determines the relative weight of squared voltage deviations versus the cost of electricity. 

Our complete formulation with the voltage in the cost function can be connected to the traditional OPF formulation that considers strict inequality constraints for voltage such as $v_{it} \leq 0.95$ or equivalently $w_{iit} \leq (0.95)^2$. In the traditional case, the Lagrangian formulation and complementary slackness imply that a penalty of the type $\lambda_{it}\max(0, (0.95)^2-w_{iit})$, where $\lambda_{it} \geq 0$ can be added to the cost function. If we assume that $\lambda_{it} = \lambda$  we obtain a penalty with constant linear growth. Instead of incorporating $\lambda_{it}$ as an optimization variable as is done in the Lagrangian formulation, we assume a single penalty $\lambda$, indicating our assumption that likely all voltages will be slack. In order to enforce the slackness, we further penalize by the squared voltage.

\section{Two-layer Coordination Architecture} \label{Architecture}

\begin{figure}[!t]
	\centering
	\includegraphics[width=2.5in]{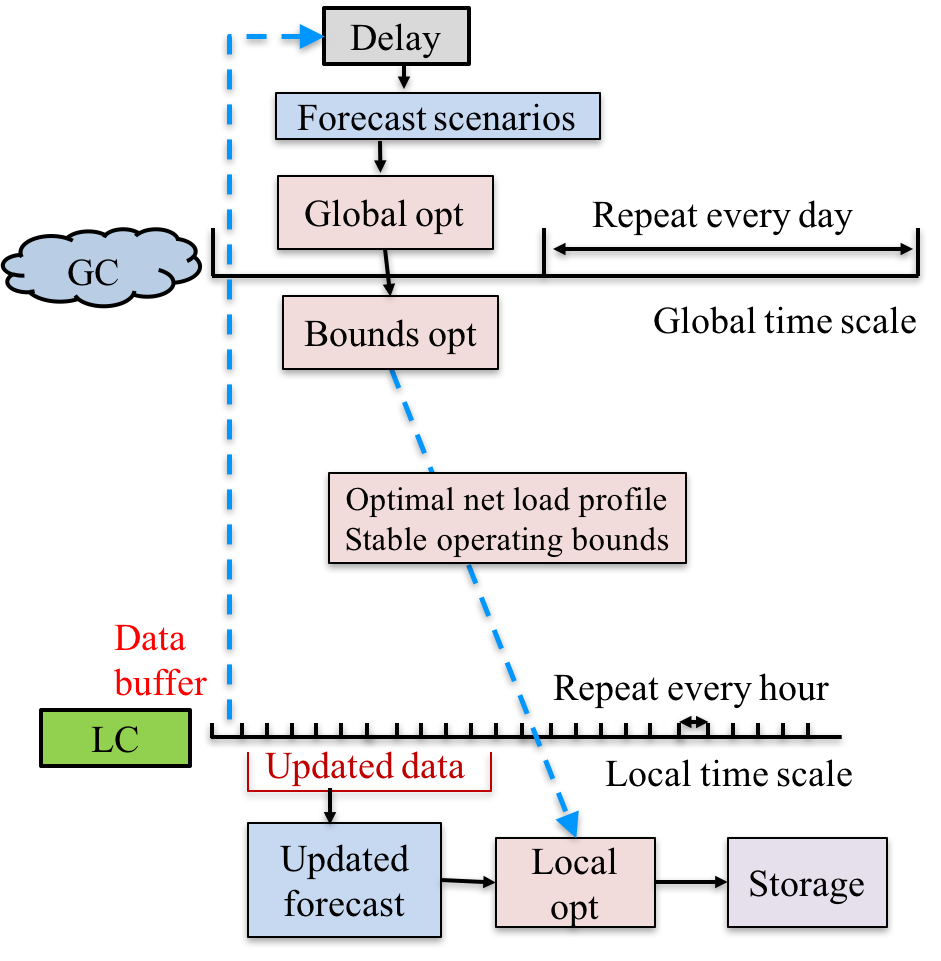}
	\caption{Flow of information in algorithm through time}
	\label{timing}
\end{figure}

The operation of the two-layer architecture is outlined in Algorithm \ref{totalAlg} and the accompanying Figure \ref{timing}. The details are discussed in the following subsections.

\begin{algorithm}[htpb]
	\caption{Process for determining storage control with two-layer architecture.}
	GC operations are repeated with period $\Delta_{\mathrm{GC}}$ which could be daily
	\begin{enumerate}
		\item GC gathers buffered net load data from nodes subject to a delay
		\item The future values of the net load are forecasted and scenarios based on these forecasts are generated
		\item GC solves a modified optimal power flow problem to determine the optimal net load profile for each LC
		\item GC solves another modified optimal power flow problem to determine the maximum and minimum feasible operating bounds for each LC
		\item The optimal net load profile and feasible operating bounds are sent to each LC
	\end{enumerate}
	LC operations are repeated with period $\delta_\mathrm{min}$, which is much smaller than $\Delta_{\mathrm{GC}}$ and could be hourly
	\begin{enumerate}
		\item LC reads its own local data in real time and maintains an updated forecast of future values
		\item LC performs a local optimization to determine the optimal storage action based on the feasible operating bounds and the net load profile given by the GC
		\item The storage system performs the charging instructions given by the LC
	\end{enumerate}
	\label{totalAlg}
\end{algorithm}

\subsection{Global Controller}

Upon receiving a new set of buffered and delayed data from the nodes (every $\Delta_{\mathrm{GC}}$ time steps), the GC operates in MPC fashion with a lookahead time of $\Delta_{\mathrm{F}}$ incorporated into the optimization horizon for a total horizon length of $\Delta_{\mathrm{GC}} + \Delta_{\mathrm{F}}$ time steps. The GC uses the buffered and delayed data to generate forecast scenarios of the future net loads for all the nodes. The GC then solves a large optimization problem for each forecast scenario to determine the net load profiles for all the nodes over the next $\Delta_{\mathrm{GC}} + \Delta_{\mathrm{F}}$ time steps. The net load profile solved for each scenario is averaged together to form a single net load profile. This single profile is used to solve two smaller optimization problems for each node to determine upper and lower bounds on its load over the same time period of $\Delta_{\mathrm{GC}} + \Delta_{\mathrm{F}}$ steps. The GC then sends to each node its load profile and bounds. We now detail these steps.

\noindent {\bf Forecasts and Scenarios.} The forecaster takes historical data of the uncontrollable demand and generation and potentially other features such as weather and outputs a predicted time series of the net load for each node before control variables are added. Scenarios are generated based on the forecast by adding a random residual to the forecast. The residual is drawn from a multivariate normal distribution with mean and standard deviation based on the empirical mean and standard deviation of historical forecast residuals. These scenarios allow the algorithm to consider a broader range of possible future trajectories, preventing it from over fitting to a single outcome. Since the forecaster only deals with uncontrollable data, it does not create a feedback loop for the system, but is simply an external input. Therefore, the use of forecasts and scenarios does not impact controller stability.

\noindent{\bf Global Optimization.} The global optimization is run with a look ahead period $\Delta_{\mathrm{F}}$ time steps added to its optimization horizon, but only the first $\Delta_{\mathrm{GC}}$ time steps are executed by the LC. For simplicity, we assume all the buffered data is received simultaneously before the GC calculation; however the problem can easily be extended to an asynchronous case for which the forecasts are only updated for the nodes that have sent new data. The objective of the optimization problem is to find the net load profile for each storage node that minimizes the total network energy cost and the voltage deviation metric defined in~\eqref{Vmetric}. The full convex optimization problem is as follows

\begin{subequations} \label{GCopt}
	\begin{alignat}{2}
	\underset{u,s,q,W}{\text{minimize:}} \quad & \sum_{\tau}   p_{\tau} \cdot \Re(s_{0\tau}) + \\ 
	& \lambda \sum_{i=0}^{N}\sum_{\tau}(\max(w_{ii\tau} - V_\mathrm{tol+}^2, 0) + \nonumber \\
	& \max(V_\mathrm{tol-}^2 - w_{ii\tau}, 0))^2 \nonumber \\
	\text{subject to:} \quad & \Re(s_{i\tau}) = \hat{d}_{i\tau} + u_{i\tau}, \hspace{3em} \forall i \in \mathcal{Z} \label{GCopt2} \\
	& \Re(s_{i\tau}) = \hat{d}_{i\tau}\hspace{6.2em}  \forall i \notin \mathcal{Z} \label{GCopt3} \\
	& q_{it} = \eta_i q_{i\tau-1} + u_{i\tau}\cdot \delta_{\mathrm{min}} \hspace{0.8em} \forall i \in \mathcal{Z} \label{GCopt4} \\
	& u^{\mathrm{min}}_{i}  \leq u_{i\tau} \leq u^{\mathrm{max}}_{i} \hspace{3.5em} \forall i \in \mathcal{Z} \label{GCopt5} \\
	& q^{\mathrm{min}}_{i}  \leq q_{i\tau} \leq q^{\mathrm{max}}_{i} \label{GCopt6} \\
	& s_{i\tau} = \sum_{j:(i,j)\in E} (w_{ij\tau} - w_{ii\tau})y_{ij}^{*} \label{GCopt7} \\
	& W\{i,j\}_\tau \succeq 0. \label{GCopt8}
	\end{alignat} 
\end{subequations}

The demand forecast scenario for node $i$ at time $\tau$ is denoted by $\hat{d}_{i\tau}$. Equations \eqref{GCopt2} and \eqref{GCopt3} represent the net load demanded by each node with and without storage. Equations \eqref{GCopt4}, \eqref{GCopt5}, and \eqref{GCopt6} represent the battery charging dynamics. Equations \eqref{GCopt7} and \eqref{GCopt8} represent the convex relaxation for AC power flow from \cite{low_exact_relaxation} and described in Section~\ref{NetworkModelSec}. Each constraint is assumed to hold for all time steps $\tau$ and all nodes $i$ unless otherwise specified. The output is the net load profile for storage nodes $\Re(s_{i\tau})$. This optimization is performed separately for each net load forecast scenario. The net load profile for each scenario is averaged together to produce a single profile that is sent to the bounds optimizer and the respective LCs.

\noindent{\bf Determining load bounds.} The purpose of this computation is to provide each local controller with upper and lower bounds on its load profile to avoid excessive voltage deviations. These are needed to provide the LC with the flexibility to operate away from the globally calculated load profile without causing an unacceptable increase in squared voltage deviations. The bounds are determined by solving two optimization problems for each node with storage as follows

\begin{subequations} \label{Boundsopt}
	\begin{alignat}{2}
	\underset{u,s,q,W}{\text{minimize:}} \quad & \Re(s_{j\tau} ) + \\
	& \lambda \sum_{i=0}^{N}\sum_{\tau}(\max(w_{ii\tau} - V_\mathrm{tol+}^2, 0) + \nonumber \\
	& \max(V_\mathrm{tol-}^2 - w_{ii\tau}, 0))^2 \nonumber \\
	\text{subject to:} \quad & \Re(s_{i\tau}) = \Re(\bar{s}_{i\tau}) \quad \forall i \ne j \label{Bopt2} \\
	& s_{i\tau} = \sum_{j:(i,j)\in E} (w_{ij\tau} - w_{ii\tau})y_{ij}^{*} \label{Bopt7} \\
	& W\{i,j\}_\tau \succeq 0. \label{Bopt8}
	\end{alignat} 
\end{subequations}

 The first optimization determines the minimum feasible net load for a particular storage node, $\Re(s_{j\tau} )$, and the second determines the maximum. In both problems all other nodes are assumed to follow the optimal net load profile given by the solution of the global optimization. This is a reasonable assumption since the nodes are expected to reasonably follow the profiles provided by the global optimization. The bounds optimizations can be performed in parallel since they are decoupled across time due to the removal of the storage constraints.

\subsection{Local Controller}
After each LC receives its load profile and bounds from the GC, it calculates forecasts and scenarios of its own load for each time step $\delta_\mathrm{min}$. This is similar to how the GC calculates forecasts; however, the LC uses the most updated data to be more accurate in the short term. The LC then uses the forecast scenarios and the most recent load profile and bounds from the GC to compute a battery charging action for the next period. 

\noindent{\bf Local Optimization.} Each LC operates in an MPC fashion, whereby only the latest control decision is executed but an additional period of $\Delta_{\mathrm{F}}$ is considered. To do so, the LC solves the following optimization problem independent of all other LCs

\begin{subequations} \label{Localopt}
	\begin{alignat}{2}
	\underset{x,q,u}{\text{minimize:}} \quad & \sum_{\tau} \sum_{g}  p_{\tau} \cdot z^g_{\tau} + \gamma ||z^g_{\tau} - x_{\tau}||_2  \\
	\text{subject to:} \quad & x^-_{\tau} \leq z^g_{\tau} \leq x^+_{\tau}  \\
	& z^g_{\tau} = \hat{d}^{g}_{\tau} + u_{\tau} \\
	& q_{\tau} = \eta_i q_{\tau-1} + u_{\tau}\cdot \delta_{\mathrm{min}}  \\
	& u^{\mathrm{min}}  \leq u_{\tau} \leq u^{\mathrm{max}}  \\
	& q^{\mathrm{min}}  \leq q_{\tau} \leq q^{\mathrm{max}}.
	\end{alignat}
\end{subequations}

In the objective, $\gamma \ge 0$ controls how strictly the local controller must follow the load profile provided by the global controller as opposed to performing its own local optimization. The local controller considers several forecast scenarios $\hat{d}^{g}_{\tau}$ in order to prevent over fitting to a single forecast. The variable $z^g_{\tau}$ is the expected net load after battery control for each scenario. The load profile $x_{\tau}$ and bounds $x^-_{\tau}$ and $x^+_{\tau}$ are provided by the GC. Finally, only the most recent value of $u_{\tau}$ is sent to the storage unit to be executed for that time step.

\section{Simulation results} \label{Results}
We evaluate the two-layer architecture detailed in the previous section using two different distribution networks; a 47 bus network adapted from \cite{low_network_model}, and the IEEE standard 123 bus feeder~\cite{IEEE_feeders}. The time resolution is set to $\delta_\mathrm{min}=1$ hour, the global buffer delay and time resolution is $\Delta_{\mathrm{GC}} = 24$ hours, and the MPC lookahead horizon is $\Delta_{\mathrm{F}} = 48$ hours. The number of global scenarios used is 24 and the number of local scenarios is 10. The power quality tuning parameter $\lambda$ is empirically tuned to 1000 with an allowed band of $V_\mathrm{tol+}=1.045$ and $V_\mathrm{tol-}=0.955$. The energy price follows a time of use model and is set to a peak price of 28 cents per KWh from 2pm to 9pm and to 20 cents per KWh for the rest of the day. This reflects the case in which consumers and DER providers have access only to the retail price structure and not to the wholesale market price.

The solar and storage penetrations are defined as percentages of the total demand in the network. The nodes that are assigned solar and storage are chosen at random from the nodes with nonzero load. The solar and storage assigned to each node is in proportion to its load relative to the total load of the chosen nodes. The storage allotment for each node sets the value of $q^{\mathrm{max}}_i$. The load for each node consists of an aggregation of homes chosen at random from a dataset of thousands of homes provided by PG\&E. The home demands are aggregated such that the peak demand matches the values specified by the network. The solar data is from NREL~\cite{NRELsolar} and was scaled to fit the solar penetration for each node. Since all nodes in the network are within close physical proximity, the same solar profile was used, but with random noise normally distributed with zero mean and standard deviation equal to 5\% of the peak generation added to account for variations. The forecaster used is an ARIMA model with form $(3, 0, 3)(3, 0, 3)_{24}$ adapted from \cite{ARIMA}.

In order to determine the extent to which the forecaster impacts the controller performance, limited simulations were performed with an artificial forecaster with adjustable accuracy. The forecaster modifies the data using the formula $\hat{d} = (1+x)d$ where $x$ is a random vector with equal length to $d$. Each element of $x$ is independently normally distributed with zero mean and a variance that grows linearly for the first 10 time periods up to a maximum of $\sigma^2$. This model was chosen to roughly mirror the forecast error published in Appendix B of~\cite{kyle}. The controller performance was simulated with $\sigma$ between $0.05$ to $0.3$. We found that performance decreased only moderately with an increase in $\sigma$. The robustness to forecast error can be attributed to the local controller MPC only executing the first time step, which has the least forecast error.

The chosen metrics we used to evaluate the results are arbitrage profit and the voltage deviation metric defined in~\eqref{Vmetric}. Arbitrage profit is defined as the profit earned from buying electricity at the low price and selling it at the peak price. The results are compared to the case in which there is no coordination and each storage unit autonomously charges at a constant rate during solar generation and discharges at another constant rate during the peak price period to optimize its arbitrage profit without regard to network reliability. Each simulation takes place over the first 30 days of August and the result presented is the average of 20 random initializations of solar and storage deployments in the network.

The simulations were run on a cluster of Intel Xeon CPU E5-2640 v3 processors, and the computation time in minutes for a representative set of simulations is shown in Table \ref{simTable}. This table also includes a breakdown of computation time in seconds of different parts of the algorithm for the 47 bus network with 15 storage nodes.

\begin{table}[h!]
	\begin{center}
		\caption{Simulation times for the two networks and several penetrations.}
		\label{simTable}
		\begin{tabular}{ccc}
			\textbf{Network} & \textbf{Storage nodes} & \textbf{Time (min)}\\
			\hline
			47 bus & 3 & 75 \\
			47 bus & 15 & 122 \\
			47 bus & 20 & 178 \\
			123 bus & 10 & 265 \\
			123 bus & 25 & 344 \\
			123 bus & 45 & 494 \\
		\end{tabular}
		\quad
		\begin{tabular}{cc}
			\textbf{Breakdown of computation time} & \textbf{Time (s)}\\
			\hline	
			Forecast & 5 \\
			GC & 107 \\
			Bounds & 126 \\
			LC & 7
		\end{tabular}
	\end{center}
\end{table}

\begin{figure}[!t]
	\centering
	\includegraphics[width=3.5in]{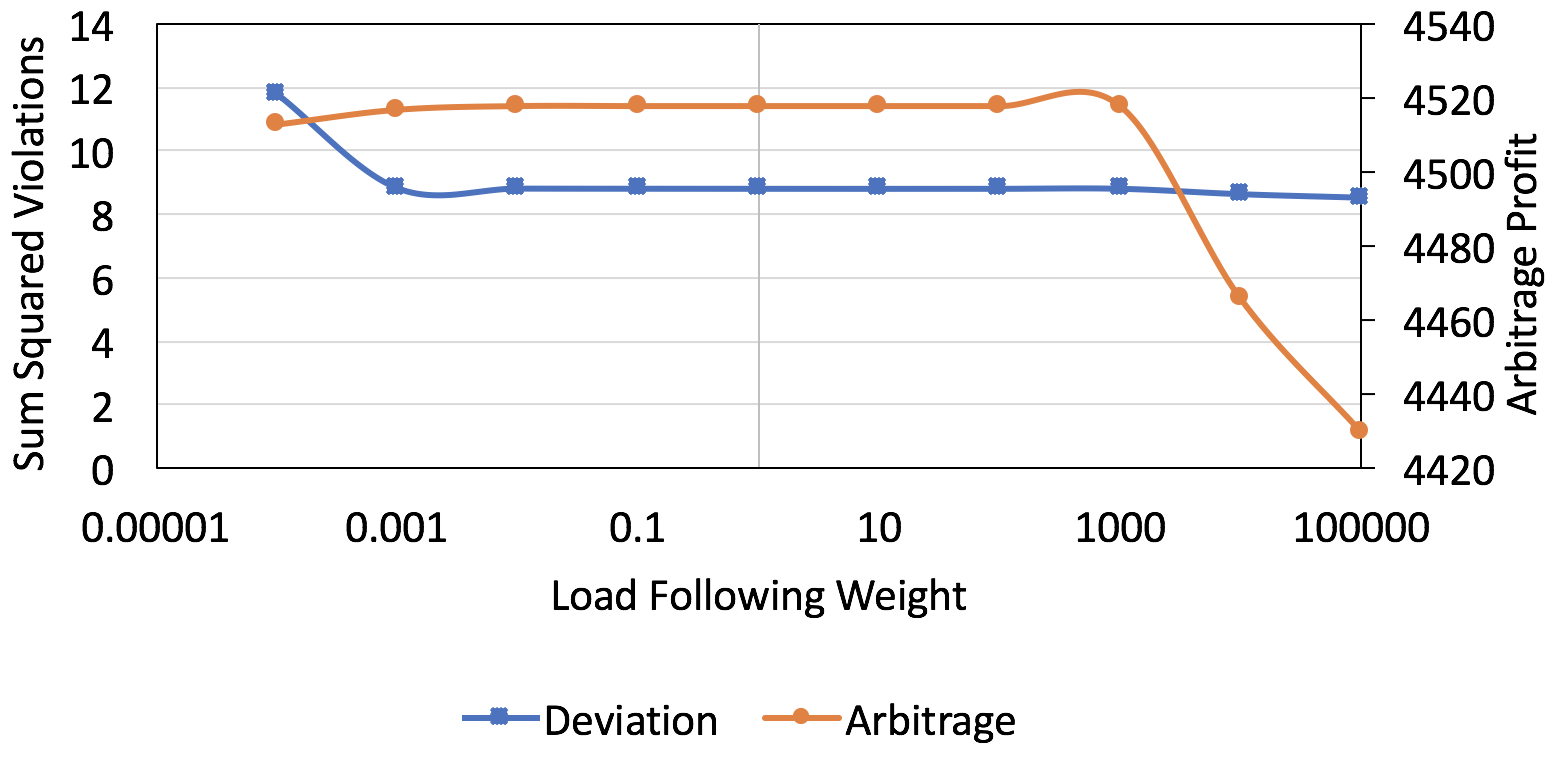}
	\caption{Arbitrage profit and squared voltage deviation vs. load following weight $\gamma$.}
	\label{NLtune}
\end{figure}

\noindent{\bf Tuning local cost savings.} The first simulations we performed are to determine the value for the local controller net load following parameter $\gamma$ on the 47 bus network. This parameter combines two control schemes explored in \cite{kyle} and is able to capture the benefits of both (high reliability and arbitrage profit). Figure \ref{NLtune} shows the results when varying $\gamma$. In the extreme case when the net load following parameter is near 0, the LC simply optimizes within the restricted bounds; however, this leads to more voltage deviations since the bounds are not perfect due to forecast error and the since the optimization is a heuristic. At the other extreme when the net load following parameter is very large approaching infinity, the LC only follows the net load from the GC without performing a local optimization. This is safer for the network, but reduces arbitrage profit. As seen, most moderate values of $\gamma$ combine the benefits of the two extreme cases and offset their shortcomings. This is because there are many feasible schedules for charging the battery that yield the same minimal cost due to the price structure, but the ones close to the global controller net load signal are able to achieve lower voltage deviations. For the following simulations, we set $\gamma = 100$.

\begin{figure}[!t]
	\centering
	\includegraphics[width=3.5in]{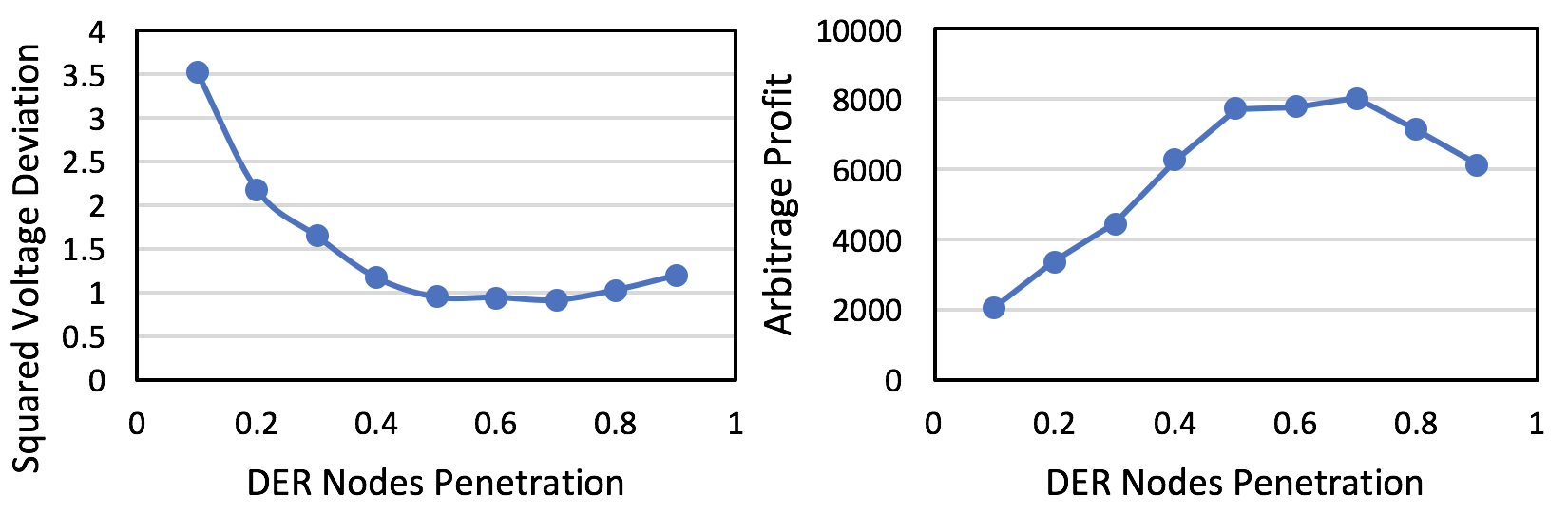}
	\caption{Arbitrage profit vs. DER nodes penetration for solar $= 0.6$ and storage $= 0.4$.}
	\label{arbspread}
\end{figure}

\noindent{\bf Effects of distributing resources.} Here we determine the effects of distributing storage and solar across the 47 bus network as opposed to allocating it among a few nodes. Figure \ref{arbspread} shows the squared voltage deviation and arbitrage profits captured by the controller when the solar and storage penetration is fixed at 0.6 and 0.4, respectively, but the percentage of nodes in the network with solar and storage is varied. As seen, when concentrating the storage within a small fraction of the nodes, the network is less capable of handling voltage deviations. Similarly, when the storage is too spread out, it is insufficient to alleviate the voltage deviations at the nodes. As seen, the network is unable to achieve maximum arbitrage profit without causing excess voltage deviations when the storage is concentrated among a few nodes. Performance is the best when the fraction of nodes with solar/storage is roughly equal to the solar penetration.

\subsection{Squared Voltage Deviation}
\begin{figure}[!t]
	\centering
	\includegraphics[width=3in]{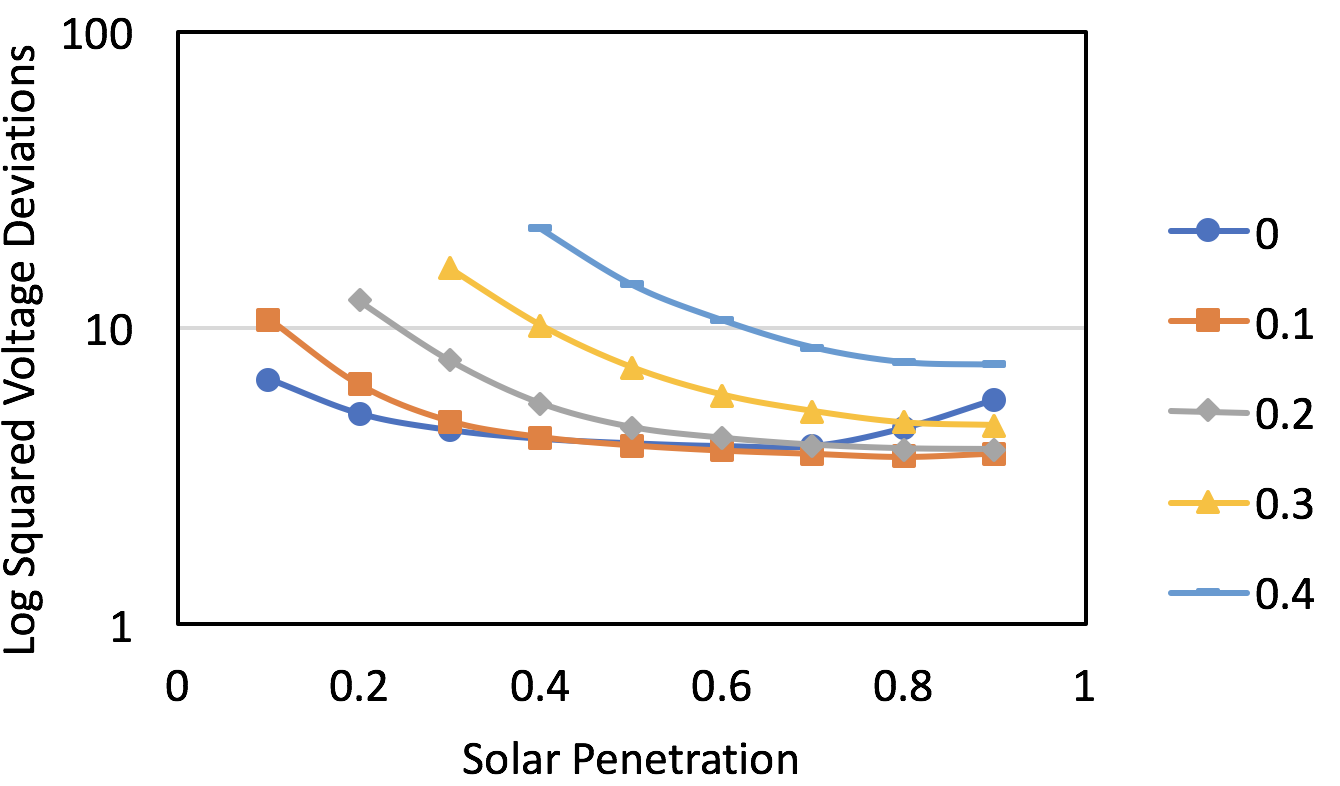}
	\caption{Log squared voltage deviation vs. solar penetration for different storage penetrations without coordination for 47-bus network.}
	\label{VnoControl}
\end{figure}

\begin{figure}[!t]
	\centering
	\includegraphics[width=3in]{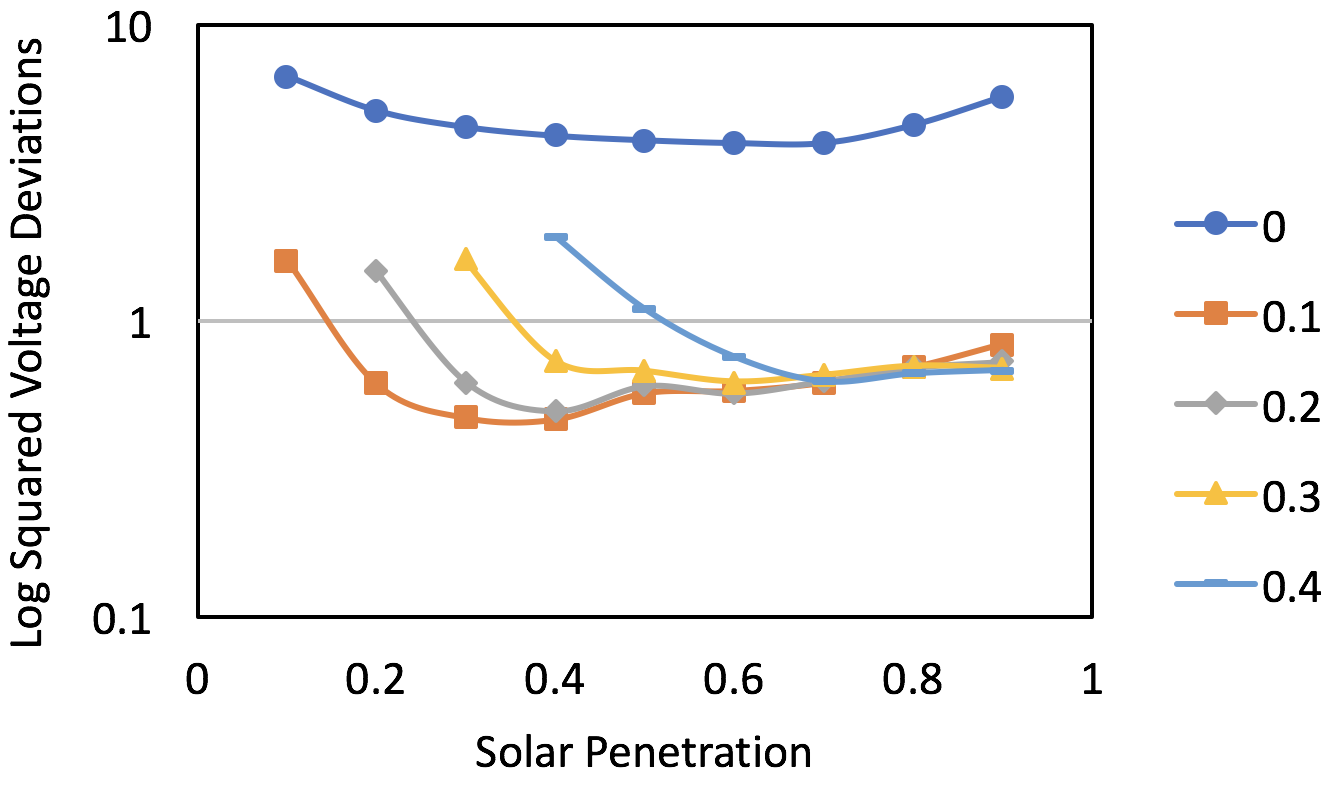}
	\caption{Log squared voltage deviation vs. solar penetration for different storage penetrations with coordination on 47-bus network.}
	\label{VControl}
\end{figure}

This simulation study quantifies the effectiveness of the DER coordination controller on the 47 bus network for different solar and storage penetrations. The performance of our control scheme is compared to the case in which storage units operate autonomously to minimize their own costs without regard to the network. Figures \ref{VnoControl} and \ref{VControl} plot log squared voltage deviation for the 47 bus network without and with coordination, respectively. For the case with no coordination, adding storage causes little to no benefit to voltage deviations as the nodes are unaware of their effects on the network. The case with no storage is the same on both graphs and is added as a reference for comparison purposes. Regardless of the storage penetration, the coordination scheme is able to make a significant improvement of over tenfold reduction in squared voltage deviation. 

\begin{figure}[!t]
	\centering
	\includegraphics[width=3in]{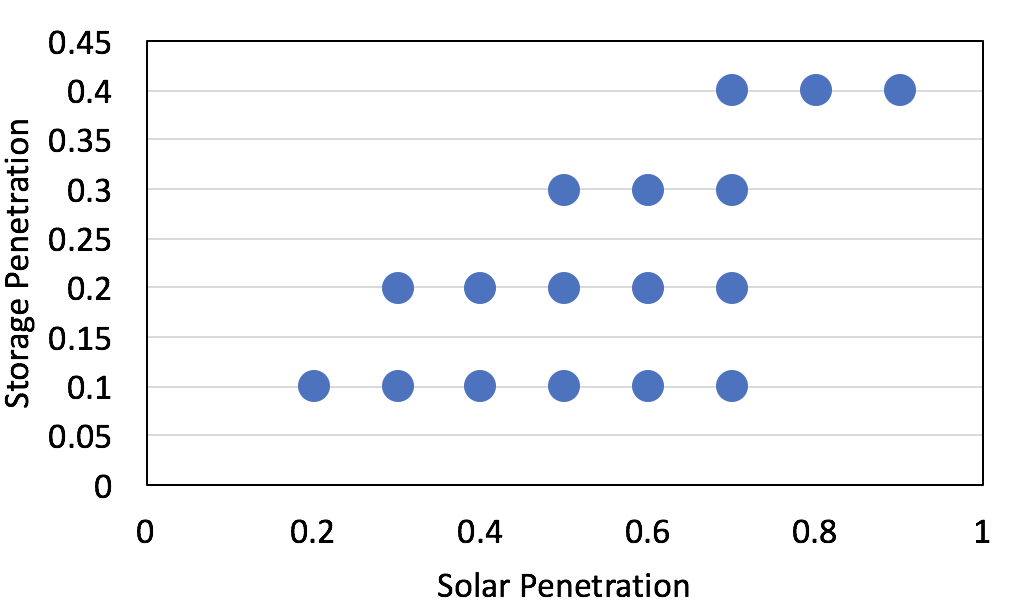}
	\caption{Set of feasible solar and storage penetrations for a squared voltage deviation below 0.68.}
	\label{FeasibleStorage}
\end{figure}

Figure \ref{FeasibleStorage} shows a set of feasible solar and storage penetrations that achieve a squared voltage deviation below 0.68, which represents a tenfold reduction compared to the baseline with no storage and 10\% solar. This clearly demonstrates the efficacy of coordination even with small amounts of storage compared to solar penetration, for example, for 70\% solar penetration only 10\% storage penetration is needed. 

\begin{figure}[!t]
	\centering
	\includegraphics[width=3in]{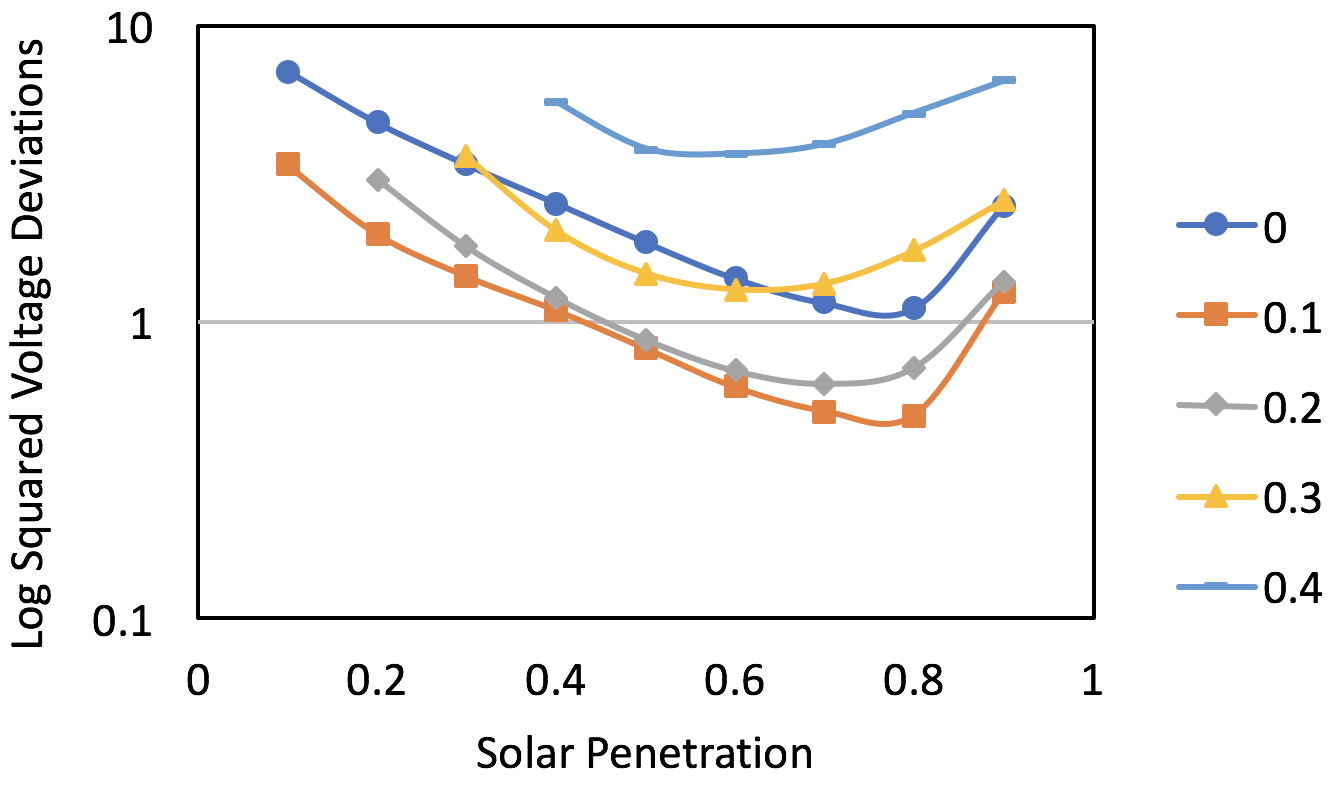}
	\caption{Log squared voltage deviation vs. solar penetration for different storage penetrations without coordination on 123-bus network.}
	\label{VnoControl123}
\end{figure}

\begin{figure}[!t]
	\centering
	\includegraphics[width=3in]{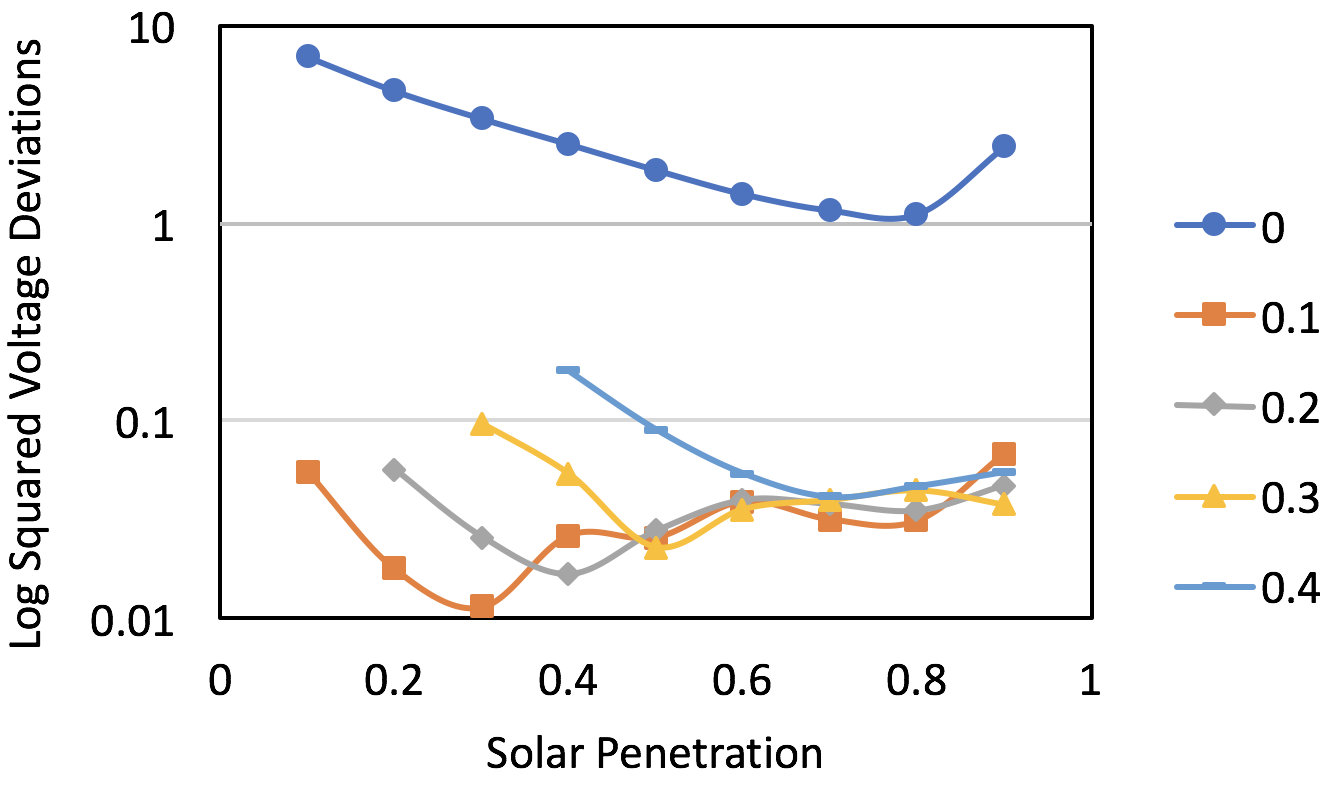}
	\caption{Log squared voltage deviation vs. solar penetration for different storage penetrations with coordination on 123-bus network.}
	\label{VControl123}
\end{figure}

As shown in Figures~\ref{VnoControl123}, ~\ref{VControl123} coordination has an even greater impact on performance for the IEEE 123-bus network.

\subsection{Arbitrage Profit}
\begin{figure}[!t]
	\centering
	\includegraphics[width=3in]{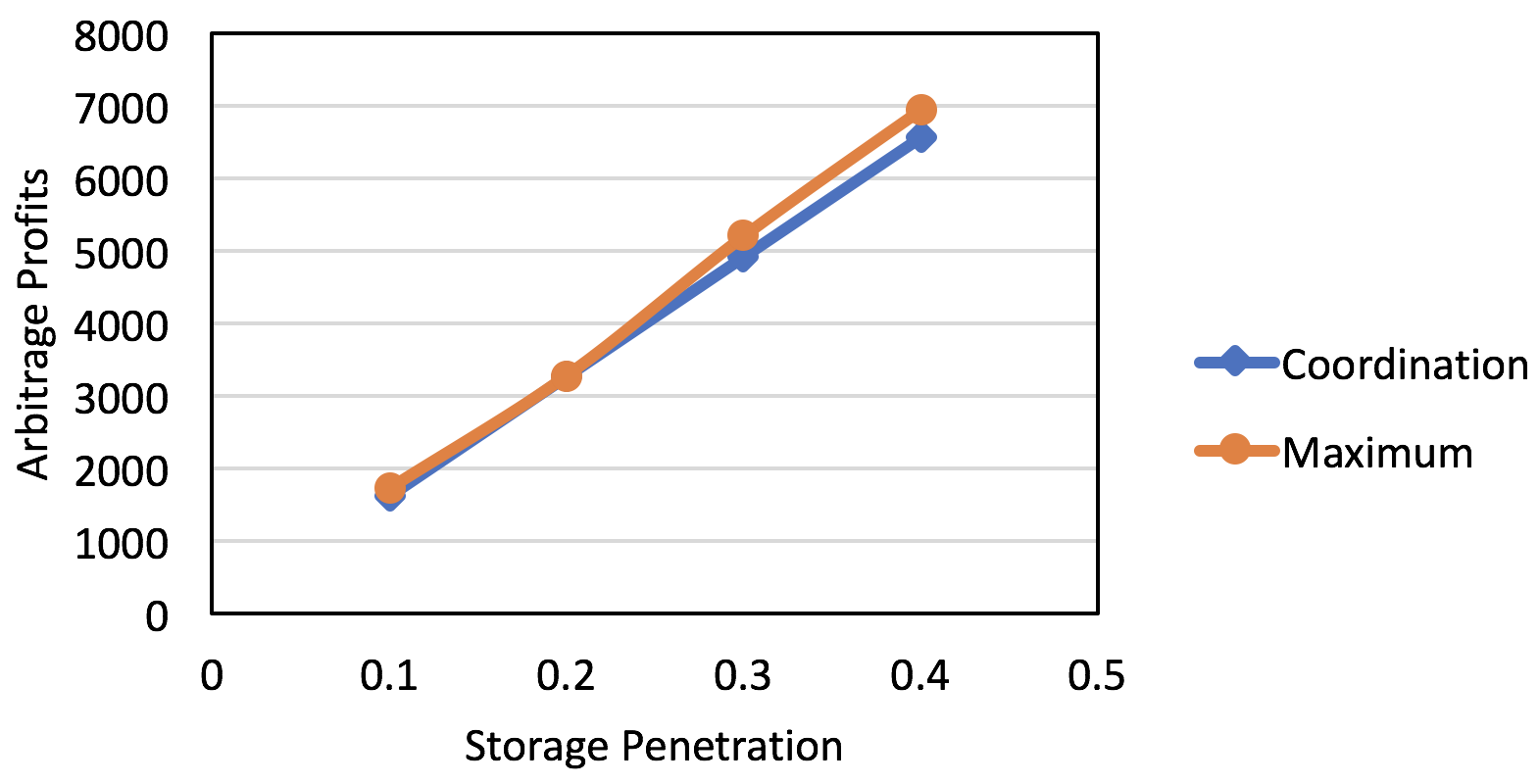}
	\caption{Arbitrage profit vs. storage penetration compared to maximum achievable.}
	\label{arbcomp}
\end{figure}

Figure \ref{arbcomp} compares arbitrage profits with and without coordination for a solar penetration of 40\%, which is representative of other solar penetrations. When there is no coordination, the maximum possible arbitrage profit is obtained since the storage units use a strategy that only maximizes profits without considering network effects. Note that the arbitrage profit with coordination is very close to the maximum. This demonstrates that the two-layer coordination architecture is able to achieve much higher network reliability with minimal loss of arbitrage profits (see Figure~\ref{arbcomp}).

\bibliographystyle{IEEEtran}
\bibliography{references_pes_v2.bib}

\end{document}